\documentclass[final,letterpaper,5p]{elsarticle}
\usepackage{lineno}
\usepackage{graphicx,color}
\usepackage{amssymb}
\usepackage{amsmath}
\usepackage{xspace}                                                 
\usepackage{latexsym}
\usepackage{amsxtra}
\usepackage{amscd}
\usepackage{dcolumn}
\usepackage{bm}
\usepackage{hyperref}
\usepackage{url}
\usepackage[T1]{fontenc} 
\usepackage{threeparttable}
\usepackage{dcolumn}
\newcolumntype{d}[1]{D{.}{.}{#1}}
\newcolumntype{k}{D{(}{(}{-3}}
\biboptions{numbers,sort&compress}
\graphicspath{{plots/}}

\newcommand{\nuc}[2]{$^{#1}$#2}
\newcommand{\gray}{$\gamma$ ray\xspace}
\newcommand{\grays}{$\gamma$ rays\xspace}
\newcommand{\ghray}{$\gamma$-ray\xspace}

\renewcommand{\ge}{$^{62}$Ge\xspace}
\newcommand{\ga}{$^{62}$Ga\xspace}
\newcommand{\zn}{$^{62}$Zn\xspace}

\usepackage{color}
\definecolor{nicegreen}{RGB}{46,204,64}

\journal{Physics Letters B}

\begin{document}
\begin{frontmatter}

\title{Isospin symmetry in the $T=1, A=62$ triplet}
\author[gsi,ut,rnc,csic]{K.~Wimmer}
\cortext[cor1]{Corresponding author}
\ead{k.wimmer@gsi.de}
\author[jyfl]{P.~Ruotsalainen}

\author[upa,infn]{S.M.~Lenzi}
\author[uam]{A.~Poves}
\author[ific]{T.~H{\"u}y{\"u}k}
\author[rnc]{F.~Browne}
\author[rnc]{P.~Doornenbal}
\author[ut,rnc]{T.~Koiwai}
\author[gsi]{T.~Arici}
\author[jyfl]{K.~Auranen}
\author[york]{M.A.~Bentley}
\author[infn]{M.~L.~Cort\'es}
\author[jyfl,orsay]{C.~Delafosse}
\author[jyfl]{T.~Eronen}
\author[jyfl,gsi]{Z.~Ge}
\author[jyfl]{T.~Grahn}
\author[jyfl]{P.~T.~Greenlees}
\author[jyfl]{A.~Illana}
\author[cns]{N.~Imai}
\author[jyfl]{H.~Joukainen}
\author[jyfl]{R.~Julin}
\author[csic]{A.~Jungclaus}
\author[jyfl]{H.~Jutila}
\author[jyfl]{A.~Kankainen}
\author[cns]{N.~Kitamura}
\author[nscl]{B.~Longfellow}
\author[jyfl]{J.~Louko}
\author[orsay]{R.~Lozeva}
\author[jyfl]{M.~Luoma}
\author[rnc]{B.~Mauss}
\author[lnl]{D.R.~Napoli}
\author[ut]{M.~Niikura}
\author[jyfl]{J.~Ojala}
\author[jyfl]{J.~Pakarinen}
\author[york]{X.~Pereira-Lopez}
\author[jyfl]{P.~Rahkila}
\author[upa,infn]{F.~Recchia}
\author[jyfl]{M.~Sandzelius}
\author[jyfl]{J.~Sar\'{e}n}
\author[ut,rnc]{R.~Taniuchi}
\author[jyfl,liverpool]{H.~Tann}
\author[york]{S.~Uthayakumaar}
\author[jyfl]{J.~Uusitalo}
\author[csic]{V.~Vaquero}
\author[york]{R.~Wadsworth}
\author[jyfl]{G.~Zimba}
\author[york,jazan]{R.~Yajzey}

\address[gsi]{GSI Helmholtzzentrum f\"{u}r Schwerionenforschung, D-64291 Darmstadt, Germany}
\address[ut]{Department of Physics, The University of Tokyo, Hongo, Bunkyo-ku, Tokyo 113-0033, Japan}
\address[rnc]{RIKEN Nishina Center, 2-1 Hirosawa, Wako, Saitama 351-0198, Japan}
\address[csic]{Instituto de Estructura de la Materia, CSIC, E-28006 Madrid, Spain}
\address[jyfl]{Accelerator Laboratory, Department of Physics, University of Jyv\"askyl\"a, FI-40014 Jyv\"askyl\"a, Finland}
\address[upa]{Dipartimento di Fisica e Astronomia dell' Universit\`a di Padova, Padova I-35131, Italy}  
\address[infn]{INFN, Sezione di Padova, Padova I-35131, Italy} 

\address[uam]{Departamento de F\'isica Te\'orica and IFT UAM-CSIC, Universidad Aut\'onoma de Madrid, 28049 Madrid, Spain}
\address[ific]{Instituto de Fisica Corpuscular, CSIC-Universidad de Valencia, E-46071 Valencia, Spain}
\address[york]{School of Physics, Engineering and Technology, University of York, YO10 5DD York, United Kingdom}
\address[lnl]{INFN, Laboratori Nazionali di Legnaro, Legnaro (Padova), Italy}
\address[cns]{Center for Nuclear Study, University of Tokyo, Hongo, Bunkyo-ku, Tokyo 113-0033, Japan}
\address[nscl]{National Superconducting Cyclotron Laboratory and Department of Physics and Astronomy, Michigan State University, East Lansing, MI 48824 USA}
\address[orsay]{Universit\'{e} Paris-Saclay, CNRS/IN2P3, IJCLab, 91405 Orsay, France}
\address[liverpool]{University of Liverpool, Liverpool L69 7ZE, United Kingdom}
\address[jazan]{Department of Physics, Faculty of Science, Jazan University, Jazan 45142, Saudi Arabia}

\begin{abstract}
Excited states in the $T_z = 0, -1$  nuclei \ga and \ge were populated in direct reactions of relativistic radioactive ion beams at the RIBF. Coincident \grays were measured with the DALI2$^+$ array and uniquely assigned to the $A=62$ isobars. In addition, \ge was also studied independently at JYFL-ACCLAB using the ${}^{24}$Mg(${}^{40}$Ca,$2n$)${}^{62}$Ge fusion-evaporation reaction. The first excited $T=1, J^\pi =2^+$ states in \ga and \ge were identified at $979(1)$ and $965(1)$~keV, respectively, resolving discrepant interpretations in the literature. States beyond the first $2^+$ state in \ge were also identified for the first time in the present work. The results are compared with shell-model calculations in the $fp$ model space. Mirror and triplet energy differences are analyzed in terms of individual charge-symmetry and charge-independence breaking contributions. The MED results confirm the shrinkage of the $p$-orbits' radii when they are occupied by at least one nucleon on average. 
\end{abstract}
\date{\today}
\end{frontmatter}

\paragraph*{Introduction}
Shortly after the discovery of the neutron, it was realized that protons and neutrons interact identically under the strong nuclear force. They can therefore be regarded as the same particle with isospin quantum number $t$ and its projection $t_z$, where $t_z=-1/2$ identifies a proton and $t_z=+1/2$ a neutron. 
This leads to the concept of charge symmetry and independence, where the strong interaction does not distinguish between proton-proton, neutron-neutron, or proton-neutron interactions. 
In nuclei, isobaric analogue states are characterized by the total isospin quantum number, $T$, with projection $T_z = \sum{t_z} = (N-Z)/2$. In the absence of electromagnetic effects, and assuming isospin symmetry, the excitation energies of these analogue states are identical. However, the electromagnetic interaction and isospin breaking terms of the nuclear interaction can lead to measurable differences in isospin multiplets. In particular, the mirror energy differences (MED), defined as $E(J^\pi,T_z=-T) - E(J^\pi,T_z=+T)$ and the triplet energy differences (TED), $E(J^\pi,T_z=-T) + E(J^\pi,T_z=+T) - 2E(J^\pi,T_z=0)$ are sensitive to the isovector and isotensor terms of the nuclear interaction, respectively~\cite{bentley07,bentley15}.

The aim of this work is to bridge the region of $N\sim Z$ nuclei from the doubly magic \nuc{56}{Ni} through the upper $fp$ shell towards the strong ground state deformation observed at $A\sim 80$~\cite{lister87}. At the start of this shell, the first spectroscopy of the $T_z=-2$ nucleus \nuc{56}{Zn} was recently published~\cite{fernandez21}. For the $A=62;\;T=1$ triplet, candidates for $J^\pi=2^+;\;T=1$ states in \nuc{62}{Ga} and \nuc{62}{Ge} have also been tentatively assigned in previous works~\cite{rudolph04,david13,henry15}.

In \nuc{62}{Ga}, $T=1$ candidates for $J^\pi = 2^+$ at 1017~keV and $4^+$ at 2234~keV were populated in a fusion-evaporation reaction and assigned indirectly based on the measured angular correlations and distributions of $\gamma$ rays and shell-model calculations~\cite{rudolph04}. 
In a similar experiment, but with higher statistics, a 979-keV transition was observed and assigned to feed the ground state~\cite{david13}. 
The angular correlation ratio for the 979-keV transition suggested a $\Delta L = 1$ dipole character, which led to a $J^\pi = 1^+$ assignment for this state since negative parity states are not expected at such a low excitation energy.
Transitions at 1017~keV and 978~keV were also observed in the $\beta$ decay of \ge and assigned to de-excite two low-lying $1^+$ states in \ga~\cite{grodner14}. However, the branching ratios from the 1017-keV state to the first excited 571-keV state and the ground state of \ga do not agree with the in-beam works of Refs.~\cite{rudolph04,david13}.
The $T=1, J^\pi = 2^+$ assignment for the 1017-keV state in Ga implies a rather large Coulomb energy difference of 63~keV and with the tentative assignment for the $2^+_1$  state in \ge, an exceptionally large TED of $-116$~keV, questioning the assignments. This led to a new experiment where \ga was populated by nucleon removal reactions from secondary \nuc{64}{Ga} and \nuc{65}{Ge} beams~\cite{henry15}. In these reactions, which are both expected to specifically populate the $T=1, J^\pi = 2^+$ state, a \gray at 977(2)~keV was observed. 
Most recently, \ga was studied again using a fusion-evaporation reaction with a \nuc{6}{Li} beam and based on the angular anisotropy, the 978~keV transition was assigned an $E2$ multipolarity~\cite{mihai22}, but the $T=1$ nature could not be proven.

In contrast to \ga, excited states in \ge have been studied previously only once. Preliminary results using a fusion-evaporation reaction are mentioned in Ref.~\cite{rudolph05}. A transition at 964~keV was tentatively assigned to \ge. While unique identification of $A$ or $Z$ was impossible, gating on \grays around 964~keV resulted in an enhancement in the energy loss spectrum of ions measured in the focal plane of the spectrometer where $Z=32$ ions are expected. No definite conclusive assignment of \grays or states in \nuc{62}{Ge} was possible.


In this letter, we present new experimental data on \ga and \ge obtained with direct nuclear reactions, inelastic scattering, one-nucleon knockout, and fusion\--\-e\-va\-po\-ra\-tion reactions. These data resolve the conflict around the $T=1, J^\pi = 2^+$ state in \ga and allow for the first unambiguous spectroscopy of the $T_z=-1$ nucleus \ge.








\paragraph*{Experimental setup and analysis of the RIBF experiment}
An experiment on \ga and \ge was performed at the Radioactive Isotope Beam Facility operated by the RIKEN Nishina Center and CNS, University of Tokyo. Proton-rich beams were produced by projectile fragmentation of a \nuc{78}{Kr} primary beam accelerated to 345~$A$MeV on a 7-mm thick primary Be target. Reaction products were separated and identified in the BigRIPS fragment separator~\cite{kubo12} by the $ToF - B\rho - \Delta E$ method through measurements of the time-of-flight, trajectory in the magnetic field, and energy loss of the ions using the standard detection systems consisting of plastic scintillators, parallel plate avalanche counters, and an ionization chamber. Data were taken in two settings - one centered on the least exotic \nuc{62}{Zn} secondary beam at 5700 pps, and the other centered on \nuc{62}{Ge}, with 290 pps, where also 1800~pps of \nuc{62}{Ga} were transmitted with approximately 40\% of the total beam intensity. The secondary beam energies were around 165 $A$MeV.
Also transmitted were the $A=63$ isotopes \nuc{63}{Ge} and \nuc{63}{Ga}, each with an intensity of about 400~pps, which allowed spectroscopy of the $A=62$ triplet by nucleon-removal reactions. The outgoing beam particles were identified in the ZeroDegree spectrometer~\cite{kubo12} using the same $ToF - B\rho - \Delta E$ method as discussed above. 
The beam then impinged on a 260-mg/cm$^2$ thick C target located at the center of the DALI2$^+$ array~\cite{takeuchi14} which consisted of 226 NaI(Tl) crystals. 
For each setting, a new energy calibration was performed during a time period when the magnetic fields of the surrounding quadrupole magnets were set to the field strengths used in that specific setting.

Transition energies and $\gamma$-ray yields were determined by fitting the Doppler-corrected spectra with simulated response functions. The GEANT4 simulation toolkit~\cite{agostinelli03} was used to produce an accurate representation of the experimental setup and the reaction properties. 
The flight time through the target amounted to about 3~ps. This was on the same order as the expected lifetime of the $2^+_1$ states in the $A=62$ nuclei. 
The lifetime of the excited states thus affected the \ghray energy and hence excitation energy determination. 
Using the Monte-Carlo simulation, a systematic uncertainty of 1~keV was determined from the analysis of the well-known decay of the $2^+$ state in \nuc{62}{Zn} with an adopted half-life of 2.93(14)~ps~\cite{pritychenko12} and a transition energy of 953.8(1)~keV~\cite{ENSDF}. 
Further uncertainties for the determination of the transition energies arise from geometrical uncertainties due to the relative positioning of the target and the DALI2$^+$ crystals. This uncertainty was estimated using known transitions in less exotic nuclei and it amounted to 2~keV.

\begin{figure}[!ht]
  \centering
\includegraphics[width=\columnwidth]{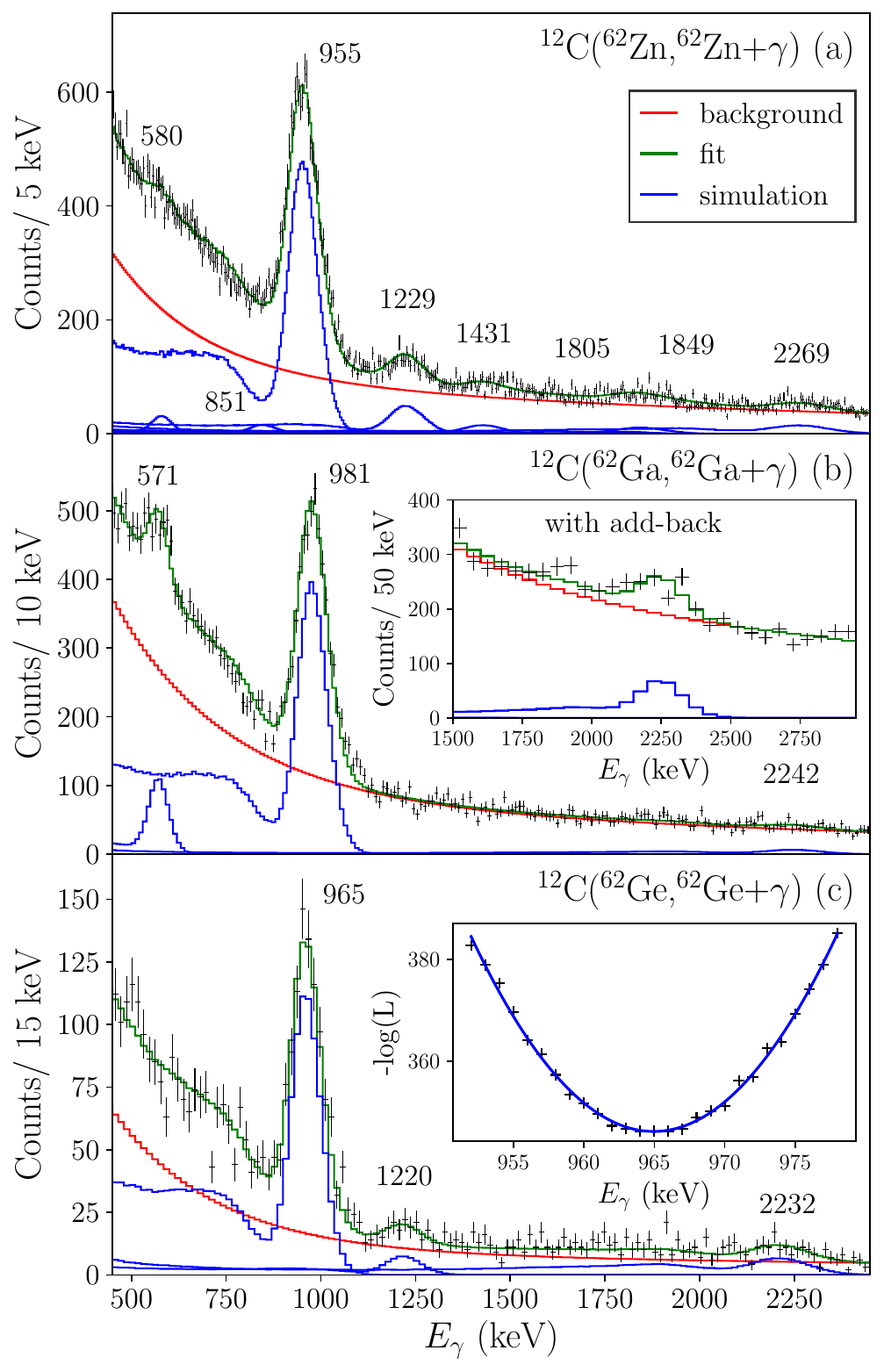}
\caption{Doppler-corrected $\gamma$-ray energy spectrum for the inelastic scattering of (a) \nuc{62}{Zn}, (b) \nuc{62}{Ga}, and \nuc{62}{Ge} on a \nuc{12}{C} target. Add-back was not applied. The peaks are labeled with their energy in keV. The Doppler correction assumes $\gamma$-ray emission at the velocity in the middle of the target and only events with a hit multiplicity of less than 5 are displayed. The data are fitted with simulated response functions for the individual transitions and a continuous background (red). For \nuc{62}{Zn}, known transitions at 851 and 580~keV have been included according to their branching ratio. The inset in panel (b) shows the high-energy region of the spectrum with add-back. Panel (c) also shows the negative log-likelihood distribution as a function of assumed transition energy for \nuc{62}{Ge}.}
\label{fig:inelastic}
\end{figure}

\paragraph*{Results of the RIKEN experiment}
The $A=62, T=1$ isospin triplet was studied by inelastic scattering as well as mirrored nucleon removal reactions from \nuc{63}{Ge} and \nuc{63}{Ga}. 
Figure~\ref{fig:inelastic} shows the Doppler-corrected spectra for the inelastic scattering of the members of the $A=62$ triplet.
In each case, the most prominent peak is associated with the decay of the first excited $T=1, J^\pi = 2^+$ state. Transition energies were determined through a maximum-likelihood fit of the measured spectrum with simulated response functions. The lifetimes extracted from the $B(E2)$ values determined in the same experiment have been used in the simulation~\cite{huyuk22}.

The known transition energies in \nuc{62}{Zn} were all reproduced within the present uncertainties which include statistical and systematic uncertainties. The energy of the first $2^+$ state was found to be $E(2^+_1) = 955(2)$~keV, in agreement with the literature value of $953.8(1)$~keV~\cite{ENSDF}. The fit of the decay of the $4^+_1$ state, $1229(4)$~keV, is in good agreement with the known adopted transition energy to the $2^+_1$ state of $1232.2(1)$~keV~\cite{ENSDF}.
In addition, higher-lying $2^+$ states at 1805 and 2803~keV were observed. Their main decay paths involve transitions of similar energies, 1805 and 1849~keV, which could not be fully resolved with DALI2$^+$. But, the observation of coincidences with the 954-keV $2^+_1 \rightarrow 0^+_1$ transition and, at the same time, an enhancement of counts in the region between 1750 and 1900~keV for hit multiplicity less than 2 events, indicates that both these states were populated in the reaction. Also, the $3^+_1$ state at 2384~keV is seen through its 1431-keV decay to the $2^+_1$ state. Weaker decay branches of the known excited states of \nuc{62}{Zn} have also been included in the fit of the spectrum with their known branching ratios. Lastly, the 2269(17)-keV transition indicated the population of the $3^-$ state at 3223.5~keV.

The highest intensity transition observed in the inelastic scattering of \nuc{62}{Ga} is located at $981(2)$~keV. This transition thus arises from the decay of the $T=1,J^\pi=2^+$ state as this is the only one expected to be this strongly populated in the inelastic scattering reaction. The cross sections for the excitation of the $2^+_1$ states in all three members of the triplet are almost identical~\cite{huyuk22}, providing additional evidence for a similar structure.
In addition, the known $1^+$ state at 571~keV was populated in the present study - the transition energy was determined to be 571(5)~keV. A transition at 2242(34)~keV was observed as well in the add-back spectra (see inset of Fig.~\ref{fig:inelastic}(b)) and found in coincidence with the $2^+_1 \rightarrow 0^+_1$ transition. The similar transition energy and excitation cross section as in \nuc{62}{Zn} suggest a $3^-$ assignment for the state at 3232(34)~keV.

The Doppler-corrected $\gamma$-ray energy spectrum for inelastic scattering of \nuc{62}{Ge} reveals three peaks. The peak at 965(3)~keV is in agreement with the tentative assignment of the $2^+_1$ state made in Ref.~\cite{rudolph05}. The other two peaks at 1220(12) and 2232(20)~keV are assigned to the decay of the newly discovered states at 2185(12) and 3197(20)~keV, respectively. Despite the low statistics, both transitions are found to be in coincidence with the $2^+_1 \rightarrow 0^+_1$ transition. The similarities in the excitation cross sections and the transition energies with \nuc{62}{Zn} suggest $4^+$ and $3^-$ assignments for these states.

\begin{figure*}[ht]
  \centering
\includegraphics[width=\textwidth]{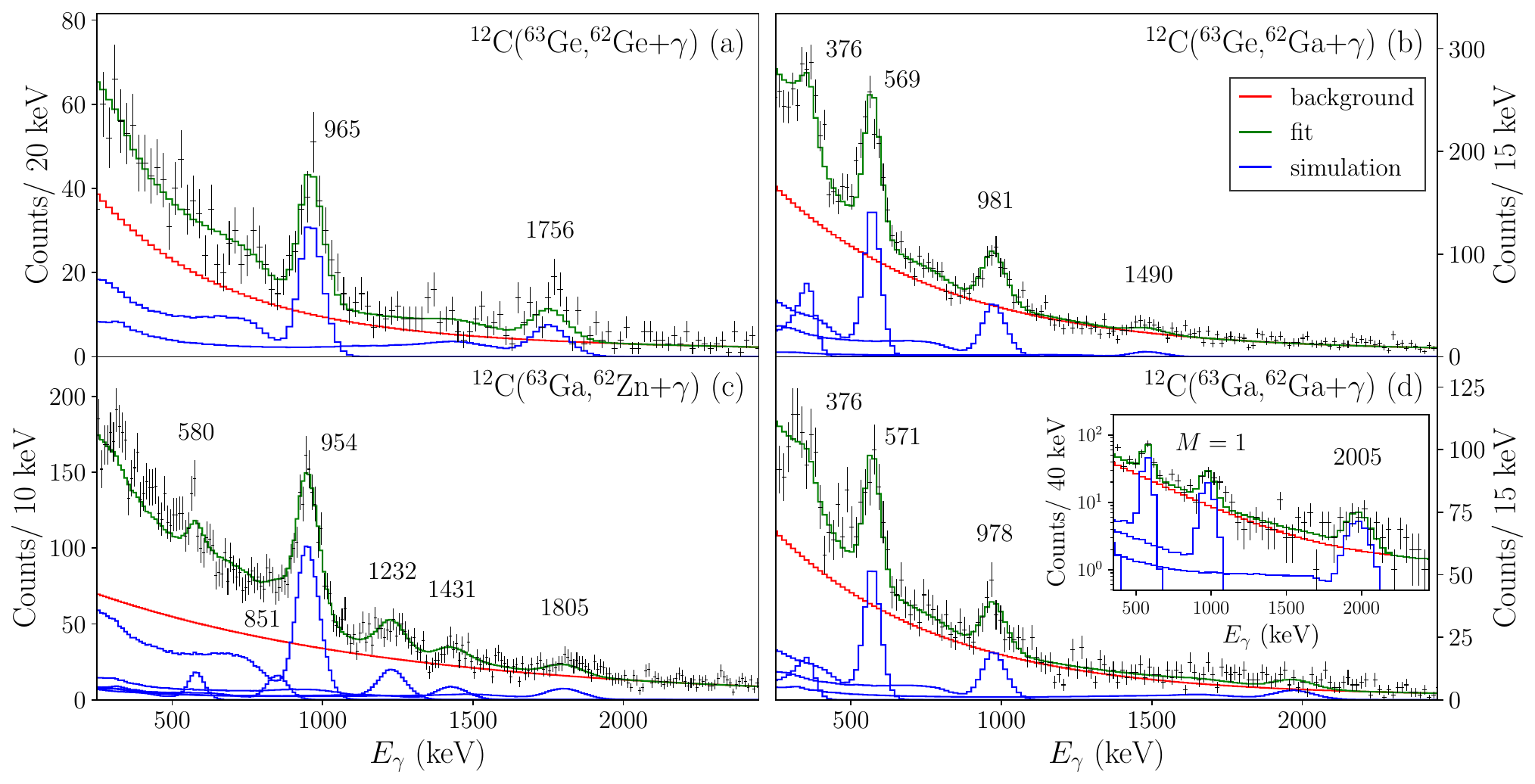}
\caption{Doppler-corrected $\gamma$-ray energy spectrum for the nucleon knockout reactions - (a) neutron and (b) proton removals from \nuc{63}{Ge}, and the mirrored proton (c) and neutron (d) removals from \nuc{63}{Ga}. The peaks are labeled with their energy in keV. The Doppler correction assumes $\gamma$-ray emission at the velocity in the middle of the target and only events with a hit multiplicity of less than 5 are displayed. The data are fitted with simulated response functions for the individual transitions and a continuous background (red). For \nuc{62}{Zn}, known transitions at 851 and 580~keV have been included according to their branching ratio. The inset in Panel (d) shows the spectrum with add-back and gated on \ghray multiplicity 1, enhancing the 2005-keV transition.}
\label{fig:knock}
\end{figure*}

The $A=62$ nuclei were also populated in the nucleon removal reactions from the \nuc{63}{Ge} and \nuc{63}{Ga} beams. Both species were present in the secondary beam as a contaminants. 
Assuming isospin symmetry, the neutron removal from \nuc{63}{Ge} and the proton removal from \nuc{63}{Ga} should lead to analogue final states in \ge and \zn.
The Doppler-corrected spectra measured in coincidence with the nucleon knockout reactions are displayed in Fig.~\ref{fig:knock}.
Panels (a) and (c) compare the analogue neutron and proton knockout reactions from \nuc{63}{Ge} and \nuc{63}{Ga}, respectively. In addition to the 965(5)~keV $2^+_1 \rightarrow 0^+_1$ transition, a previously unobserved transition at 1756(13)~keV is observed in \ge. The peak is strong in spectra gated on \ghray multiplicity 1 and no coincidences have been observed. Therefore, this transition is tentatively assigned as the $2^+_2 \rightarrow 0^+_1$ ground-state transition leading to a new state at 1756(13)~keV, which is very similar to the analogue $2^+_2$ state in \zn at 1804.7(1)~keV~\cite{ENSDF}. A decay to the first excited state, which would be expected at a transition energy of 791(14)~keV is not observed, and due to the location close to the Compton edge of the much more intense 965(5)-keV transition, no upper limit could be determined. 
%
Comparing the analogue \nuc{63}{Ge}~$\rightarrow$~\nuc{62}{Ge} and \nuc{63}{Ga}~$\rightarrow$~\nuc{62}{Zn} reactions, a similar population pattern for the ground, $2^+_1$, and $2^+_2$ states in both final nuclei is found. In the proton knockout reaction to \nuc{62}{Zn} also the $3^+_1$ and $4^+_1$ states are populated with an intensity similar to the $2^+_2$ state. This is not the case for the neutron knockout from \nuc{63}{Ge}. Based on the statistics observed in Fig.~\ref{fig:knock}\,(a), indications for the $3^+_1$ and $4^+_1$ states in \ge should have been observed. An explanation might be that these states are populated in \nuc{62}{Zn} indirectly through the decay of the higher-lying states, which in the case of \nuc{62}{Ge} are particle unbound. Population of a $4^+$ state by knockout from the $3/2^-$ ground state of the projectile requires the removal of a particle from the $0f_{7/2}$ or $0f_{5/2}$ orbital. The former is strongly bound below the $N/Z=28$ shell closure, while for the latter the shell-model calculations (see below) predict only very weak occupations. A strong direct population of $J\geq 4$ states is thus not expected.

The mirrored nucleon knockout reactions from \nuc{63}{Ge} and \nuc{63}{Ga} are expected to populate identical states in \nuc{62}{Ga}. The two spectra are shown in Fig.~\ref{fig:knock}\,(b) and (d). In both cases, strong population of the $1^+$ state at 571.2(1)~keV~\cite{ENSDF} and the $T=1,J^\pi = 2^+$ state is observed. In the proton removal reaction a previously unknown transition at 1490(20)~keV is revealed. This transition is not in coincidence with the other two lines, suggesting a decay to the ground state and a new state at 1490(20)~keV.
This state is not observed in the neutron knockout reaction. Instead, a candidate transition at $2005(54)$~keV is visible when applying add-back and gating on multiplicity 1 events in order to suppress background. This result is, however, not fully conclusive and there might be two transitions in the range of 1900-2100~keV.
In addition, the two spectra for \nuc{62}{Ga}, Fig.~\ref{fig:knock}(b,d), show an enhancement of counts around 350~keV, which is absent in the other two nuclei. These counts can be attributed to the 376.3(1)-keV decay of the $5^+_1$ state at 1193.5(2)~keV to the $3^+$ state at 817.2(1)~keV~\cite{ENSDF}. The latter state is long-lived with $\tau = 4.9(14)$~ns~\cite{rudolph04} and therefore its decay is not observed in the present experiment. A strong population of the $T=0,J^\pi = 5^+$ state was also observed in the two-neutron knockout reaction~\cite{henry15}.
The lifetime of the $5^+$ state is not known and shell-model calculations predict a lifetime beyond the sensitivity of the present experiment. Therefore, in the analysis $\tau=1$~ns was assumed.


\paragraph*{Spectroscopy of  \ge at JYFL-ACCLAB}
Shortly after the RIKEN experiment the $^{62}$Ge nucleus was studied at the Accelerator Laboratory of the University of Jyv\"{a}skyl\"{a} (JYFL-ACCLAB) employing the ${}^{24}$Mg(${}^{40}$Ca,$2n$)${}^{62}$Ge fusion\--\-e\-va\-po\-ra\-tion reaction. The ${}^{40}$Ca beam, accelerated with the K130 cyclotron to the middle-of-target energy of 106 MeV, was fused with the ${}^{24}$Mg target atoms for 244 hours with an average intensity of 3.5~pnA.
 

Prompt $\gamma$ rays were detected at the target position with the JUROGAM~3 germanium-detector array~\cite{pakarinen20}. The target position was additionally surrounded by the JYTube scintillator detector array to veto reaction channels associated with charged-particle evaporation. The fusion-evaporation recoils were further separated from the unreacted beam and other reaction products with the vacuum-mode mass-separator MARA~\cite{uusitalo19}, which was tuned to pass mass $A=62$ recoils to its focal plane. The MARA focal plane setup consisted of a multi-wire proportional counter (MWPC) and a double-sided silicon strip detector (DSSSD) to measure the recoil position, $\Delta E$, and time-of-flight. The DSSSD was also used to detect the recoil implantation and the subsequent $\beta$ decay within the same detector pixel, which allowed for recoil-decay correlations. A plastic scintillator Tuike~\cite{joukainen22} was used to measure the full remaining energy of the emitted $\beta$ particles after passing through the DSSSD. 

The ${}^{62}$Ge and ${}^{62}$Ga reaction products were identified at the MARA focal plane based on their characteristic $\beta$-decay properties. The relatively short $\beta$-decay half-lives and high $\beta$-decay endpoint energies of ${}^{62}$Ge ($T_{1/2}$ $=$ 73.5(1) ms, $Q_{EC}$~=~9730(140)~MeV~\cite{orrigo21}) and ${}^{62}$Ga ($T_{1/2}$ $=$ 116.121 (21) ms~\cite{grinyer08}, $Q_{EC}$~=~9181.1(4)~MeV~\cite{eronen06}) allowed the application of the recoil-$\beta$ tagging method~\cite{ruotsalainen13,henderson13} to identify the prompt $\gamma$ rays originating from these nuclei. 

\begin{figure}[!ht]
  \centering
\includegraphics[width=\columnwidth]{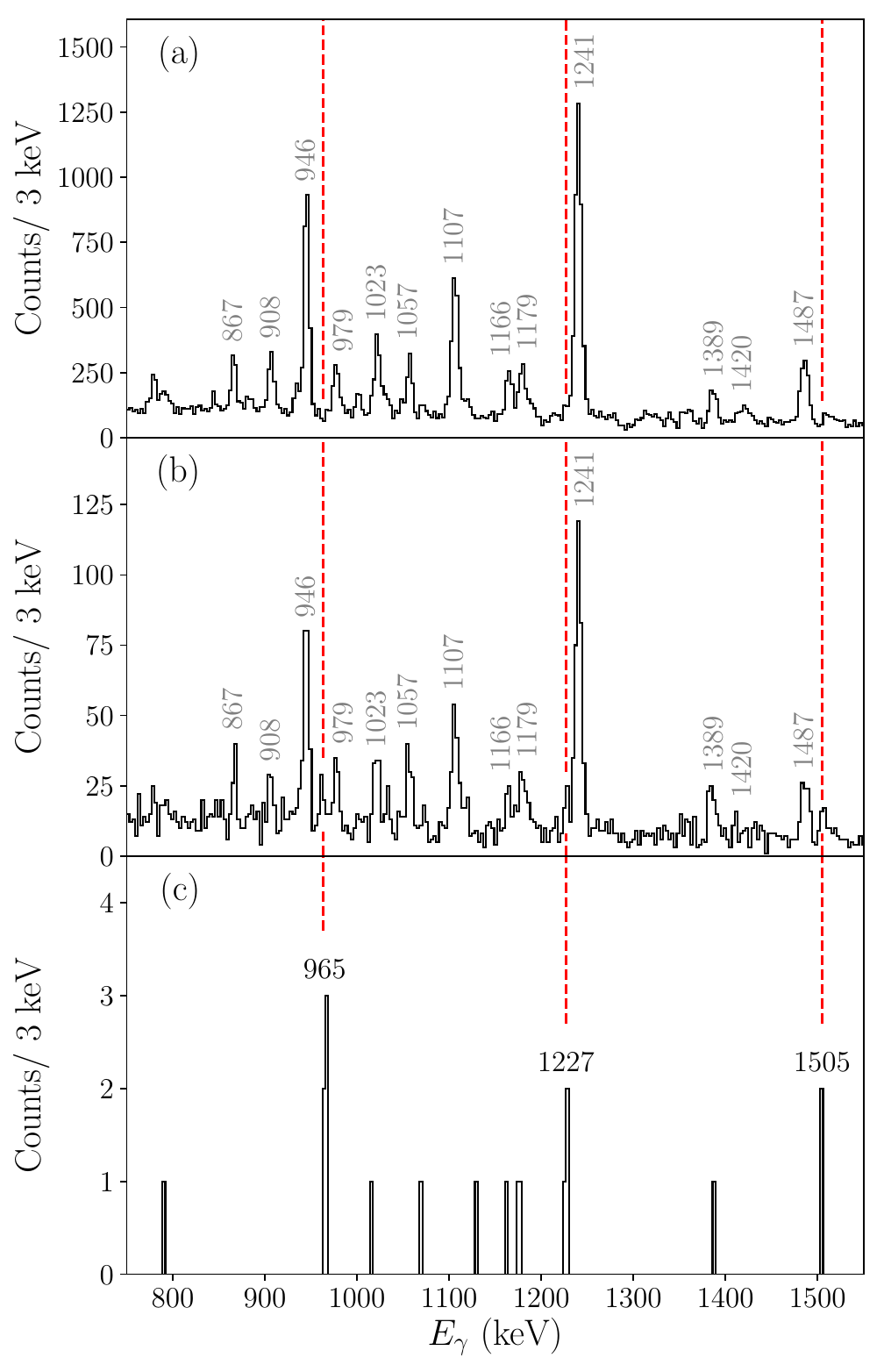}
\caption{(a) Recoil-$\beta$ correlated $\gamma$-ray spectrum with one detected proton from the $^{40}$Ca+$^{24}$Mg fusion-evaporation experiment performed at JYFL-ACCLAB, (b) same as (a), but with the requirement of zero detected protons and (c) recoil-$\beta$-$\beta$ correlated $\gamma$-rays with zero protons leading to the identification of $^{62}$Ge $\gamma$ rays. All $\gamma$-ray lines labeled in panels (a) and (b) are known transitions in \nuc{62}{Ga}. See text for further details.}
\label{fig:rbt}
\end{figure}

Since the $\beta$-decay properties of ${}^{62}$Ge (2$n$ channel) and ${}^{62}$Ga ($pn$ channel) are very similar, the JYTube charged-particle veto detector was used to differentiate between these evaporation channels. Figure~\ref{fig:rbt}\,(a) shows a recoil-$\beta$ tagged JUROGAM~3 $\gamma$-ray spectrum with recoil-$\beta$ correlation search time of 250~ms, $\beta$-particle energy threshold of 4.5~MeV and with one detected proton in JYTube. The resulting $\gamma$-ray spectrum shows known ${}^{62}$Ga transitions labeled with their energies~\cite{david13}. In Fig.~\ref{fig:rbt}\,(b) the same correlation conditions have been used as in panel (a), but now with detection of zero protons in JYTube. Since the JYTube detection efficiency for one proton was approximately 70\%, the ${}^{62}$Ga $\gamma$-ray lines are still visible in the charged-particle vetoed spectrum of Fig.~\ref{fig:rbt}\,(b). However, as a result of the charged-particle veto, three $\gamma$-ray peaks marked with the red dashed vertical lines become visible in Fig.~\ref{fig:rbt}\,(b). These are located at the energies of 965(1), 1227(2) and 1505(2)~keV and are the candidates for the \textit{T}~=~1, $2^+_1 \rightarrow 0^+_1$, $4^+_1 \rightarrow 2^+_1$ and $6^+_1 \rightarrow 4^+_1$ yrast transitions in ${}^{62}$Ge. 
To unambiguously identify the ${}^{62}$Ge $\gamma$-ray transitions, the newly established recoil-double-$\beta$ tagging (RDBT) me\-thod can be employed. The RDBT method makes use of the detection of two fast and high-energy $\beta$ particles following the recoil implantation in a single pixel. The 73.5(1)-ms, \nuc{62}{Ge} $\rightarrow$ \nuc{62}{Ga} $\beta$ decay, followed by the 116.121(21)-ms, \nuc{62}{Ga} $\rightarrow$ \nuc{62}{Zn} $\beta$ decay provide a clean tag for the ${}^{62}$Ge $\gamma$ rays. This method has been applied in Fig.~\ref{fig:rbt}\,(c) where, in addition to the charged-particle veto, the correlation search times and $\beta$-energy thresholds were 165~ms and 2.5~MeV for the first $\beta$ decay and 415~ms and 2.5~MeV for the second $\beta$ decay, respectively. As can be seen in Fig.~\ref{fig:rbt}\,(c), the RDBT method yields the same $\gamma$-ray lines as identified in the panel (b). Moreover, the 965(1)- and 1227(2)-keV $\gamma$ rays were found to be in coincidence.
\begin{figure*}[hbt]
  \centering
\includegraphics[width=\textwidth]{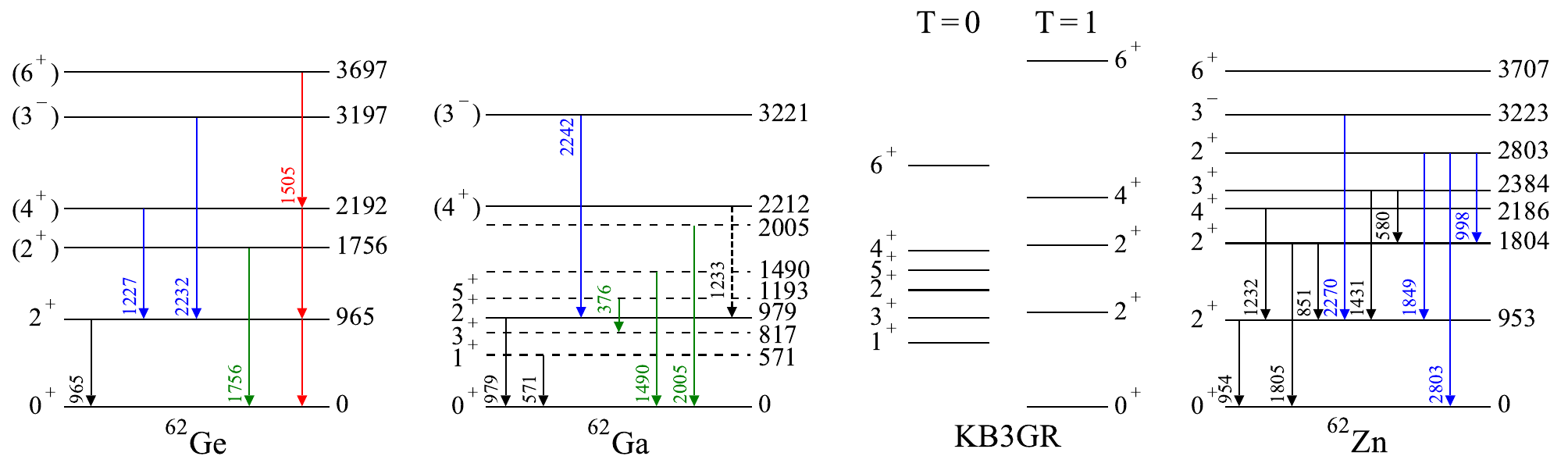}
\caption{The level schemes of \ge, \ga, and \zn determined in the present work. Level energies and transitions are labeled with their adopted energies in keV. Transitions which are only observed in the inelastic scattering are shown in blue, the ones only seen in the knockout reaction channels in green, while the ones seen in both types of reactions are marked black. The transitions observed in the JYFL-ACCLAB experiment are marked in red. Note that the 1233~keV transition in \ga was not observed in the present work and has only been placed through systematics. Also shown are the calculations for \ga employing the KB3GR interaction and a truncation at $t=8$ for the lowest $T=0$ and $T=1$ states.}
\label{fig:level}
\end{figure*}
The observed transition energies are in good agreement, within the experimental errors, with those identified in the RIKEN experiment. The fact that the same transitions for ${}^{62}$Ge are identified in two different experiments, which have been analyzed independently, gives these findings very high confidence.





\paragraph*{Discussion}
In the following discussion, we adopt the \ghray transition energies and excitation energies determined in the JYFL-ACCLAB work when available as they are more precise.
The level schemes of \ge, \ga, and \zn together with the observed \ghray transitions are presented in Fig.~\ref{fig:level}.
This work demonstrates the first unambiguous identification of the $T=1,J^\pi=2^+$ states in \ga and \ge. The strong inelastic excitation of the 979-keV state from the ground state in \ga, as well as its Coulomb excitation~\cite{huyuk22}, provide evidence for the $T=1$ quadrupole excitation. A state with very similar excitation energy was first observed in a fusion-evaporation reaction and was assigned to be $1^+$ state based on the angular correlation of \grays, albeit with large uncertainty~\cite{david13}. A peak at 979(1)~keV can be also seen in Fig.~\ref{fig:rbt}\,(a) and (b), which show the new fusion-evaporation data from JYFL-ACCLAB. Regardless, the $1^+$ interpretation was questioned as it leads to an unusual TED value~\cite{henry15} and, instead, the 979-keV state was proposed as the $T=1, J^\pi=2^+$ state. The present result confirms this interpretation and puts it on a firm footing. A 1233-keV transition, that was observed in the fusion-evaporation study in coincidence with the 979-keV transition~\cite{david13}, is a natural candidate for the $4^+_1 \rightarrow 2^+_1$ transition in \ga. In the RIKEN experiment, this transition was not observed. The $4^+_1 \rightarrow 2^+_1$ assignment for the 1233-keV transition is very tentative and previously other states have been associated with the $T=1,J^\pi=4^+$ state`\cite{rudolph04}.


For \ge, the present measurements confirmed the inferred \ghray transition at 964~keV in Ref.~\cite{rudolph05}. In addition, the observation of the 1220(12)~keV transition in the inelastic scattering of \ge on carbon and the 1227(2)-keV transition identified in the fusion-evaporation reaction allows tentative $(4^+)$ assignment for the state at 2192~keV. This is about 100~keV lower than the speculative $4^+$ assignment made in Refs.~\cite{rudolph05,johansson04} for the state at 2285~keV.
The additional transition observed at 1765~keV leads to a candidate for a $2^+_2$ state at this energy, based on comparison with the mirror nucleus \zn and the nucleon knockout cross sections.

Next, comparison of the experimental level schemes to the isospin symmetric shell-model calculations is carried out. As shown in Ref.~\cite{srivastava15}, the excitation energies calculated with the jj44b effective interaction~\cite{cheal10} for the $T=1$ states are in agreement with the experimental data for \zn.
However, it fails in the prediction of the $T=0$ states. Indeed, the adopted model space, that does not include the $f_{7/2}$ shell, is not the appropriate one for the description of the states in the $A=62$ nuclei observed in the present work. Therefore, shell-model calculations in the $fp$ model space have been performed. Up to $t=8$ nucleons were allowed to be excited across the $N=Z=28$ shell gap. The results obtained with the KB3GR interaction~\cite{poves01,kb3gr} are shown in Fig.~\ref{fig:level}.
%
The calculated excitation energies agree well with the experimental results presented in this work for \ga and \ge.
The two new states in \ga observed in the knockout reaction channels, at 1490 and 2005~keV, are likely $T=0$ states. 


It is useful to compare the level schemes of the three isobars computing the difference in excitation energy between the analogue states, the mirror and triplet energy differences.

\begin{figure}[ht]
  \centering
\includegraphics[width=\columnwidth]{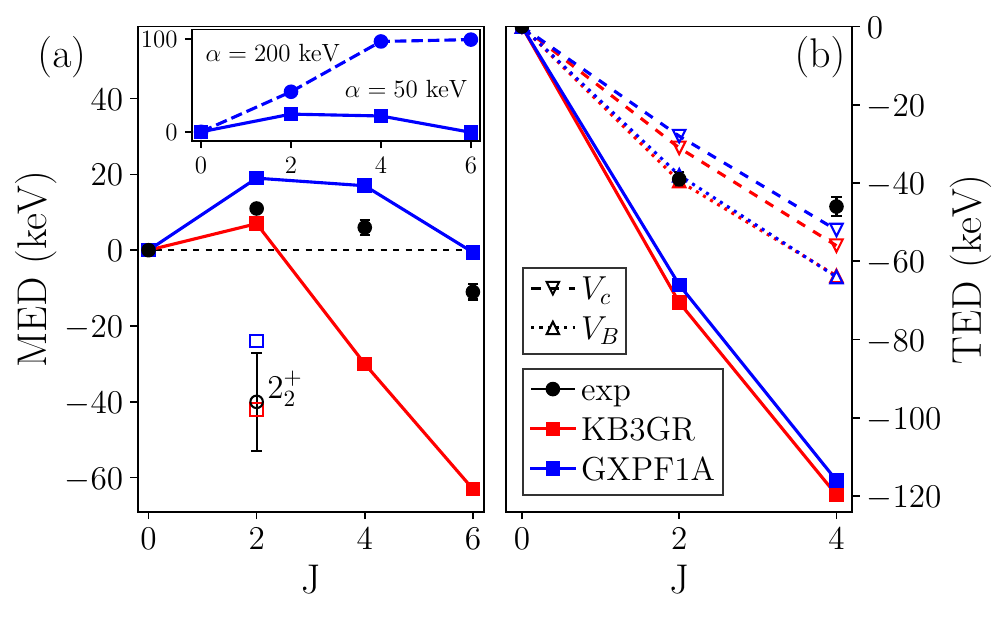}
\caption{(a) Mirror energy differences for the \ge$-$\zn pair as a function of the spin of the state. The yrare $2^+_2$ state is shown by an open symbol. The red (blue) lines show results of the shell-model calculations with the K3BGR (GXPF1A) effective interactions~\cite{poves01,honma04}. The inset shows the results for the GXPF1A calculation comparing $\alpha=50$~keV adopted here with $\alpha =200$~keV. (b) Triplet energy differences for the $A=62$ nuclei. The value at $J=4$ uses the tentative $(4^+)$ assignment in \ga based on the transition observed in Ref.~\cite{david13}. Also shown are the individual contributions to the TED.}
\label{fig:mirror}
\end{figure}

The experimental values are compared in Fig.~\ref{fig:mirror} to the shell-model calculations. The well established method developed in Refs.~\cite{lenzi01,zuker02,bentley07,bentley15} has been applied in Fig.~\ref{fig:mirror} to compute MED and TED. To obtain the MED, the Schr\"odinger equation is solved in the full $fp$ valence space using two different isospin-conserving effective interactions, KB3GR~\cite{kb3gr} and GXPF1A~\cite{honma04}, and the Coulomb and other isospin breaking interactions are treated perturbatively. Most of the Coulomb effects are taken into account by calculating the expectation value of the Coulomb interaction in the valence space $V_{CM}$, with the single-particle energy corrections for protons and neutrons given by the electromagnetic spin-orbit interaction and the orbital term~\cite{bentley07}. Another contribution to the MED arises from changes in the nuclear radius as a function of the nuclear spin. Indeed, the radius depends on the orbits that are occupied and the occupations may change for different states and hence as a function of the spin. In the $fp$ shell, $p$ orbits have larger radius than the $f$ ones, thus feeling less Coulomb repulsion. Following Refs.~\cite{lenzi01,bentley07}, we calculate the radial Coulomb contribution $V_{Cr}$ obtained from the difference of the average occupation numbers of protons ($z_p$) and neutrons ($n_p$) in the $p$ orbits at each excited state $J$ with respect to the ground state,  $V_{Cr}(J) = 2~T~\alpha [(z_p(0^+_\text{gs}) + n_p(0^+_\text{gs})) - (z_p(J) + n_p(J))]/2$, where $T$ is the isospin. The parameter $\alpha$ has been fixed to 200~keV for lighter nuclei with $N/Z=20-28$, where $p$ orbits are fractionally occupied. In a recent study~\cite{fernandez21}, it has been shown, following Ref.~\cite{bonnard18}, that this parameter has to be reduced when these low-$\ell$ orbitals are occupied by at least one particle on average. As this is the case for both $p$ orbits in the $A=62$ mirror pair, we adopt the value of $\alpha=50$~keV used in Ref.~\cite{fernandez21}. Aside from the Coulomb interaction, an additional isospin-symmetry breaking contribution $V_B$ has been identified from the systematic analysis of MED in the $f_{7/2}$ shell~\cite{zuker02}. Following Ref.~\cite{bentley15}, it is calculated using a schematic isovector interaction that introduces a $-70$-keV difference between the matrix elements of two protons minus two neutrons coupled to angular momentum zero.
The MED are quite sensitive to the nuclear structure and therefore constitute a stringent test to the effective interactions. In the present case, the yrast MED are unusually small, and in the theoretical calculations they result from a partial cancellation of the multipole terms and a radial term.
They are rather well reproduced by shell-model calculations using two different effective interactions, with KB3GR being closer to data at low spin and GXPF1A at higher spins. From the analysis of the wave function composition, it results that calculations using the KB3GR effective interaction favor more excitations from the $f_{7/2}$ shell to the upper orbitals for both protons and neutrons than those using  GXPF1A. In particular, the wave functions obtained with KB3GR for the $T=1$ states feature one proton and one neutron particle-hole excitations across the $Z=N=28$ shell gap, while for GXPF1A this constitutes less than 30\% of the wave functions.
 
The TED show a typical negative trend as a function of $J$.
This behavior is observed for all measured triplets~\cite{lenzi18} so far. This can be explained in part by Coulomb effects in terms of nucleon re-coupling~\cite{lenzi99}, but other isospin-breaking effects have to be considered as well~\cite{zuker02} to reproduce the experimental data in the shell-model framework. A similar procedure to that for MED is used to compute TED. However, due to the way TED are defined, monopole terms cancel out and only the multipole Coulomb and the $V_B$ terms contribute. For the latter, following ~\cite{zuker02,lenzi18} we use an isotensor term with a strength of 100~keV in the zero-coupling channel. As for other $T=1$ triplets, the $V_{CM}$ and $V_B$  contributions result in nearly equal strengths using both effective interactions, as shown in Fig~\ref{fig:mirror}\,(b). However, when summed together they overestimate the data. Similar results are obtained in Ref.~\cite{mihai22} by means of beyond-mean-field  calculations using realistic interactions.

In summary, excited states in the members of the $A=62;\;T=1$ isospin triplet have been populated and identified in two different experiments. The combination of the two data sets from direct inelastic scattering and nucleon removal reactions as well as fusion-evaporation reactions enabled the unique identification of the $J^\pi=2;\;T=1$ states in \nuc{62}{Ga} and \nuc{62}{Ge}. This and further spectroscopy of the yrast states allows for comparison of analogue states in the isobaric triplet for the first time. 
Both the measured mirror and triplet energy differences are rather small. The shell-model analysis shows that the MED result from the partial cancellation of the multipole $V_{CM}$ and $V_B$ terms, and a marginal contribution of the monopole radial term $V_{Cm}$, confirming the need to take into account the reduction of the radii of the $p$ orbits when they are occupied in average by more than one nucleon~\cite{fernandez21, bonnard18}. 


\paragraph*{Acknowledgments}
We would like to thank the RIKEN accelerator, the BigRIPS, and the JYFL-ACCLAB teams for providing the high intensity beams. Dr. Bettina Lommel from the GSI target laboratory is acknowledged for providing the enriched ${}^{24}$Mg target for the JYFL-ACCLAB experiment.

K. W. acknowledges the support from the Spanish Ministerio de Ciencia, Innovaci\'on y Universidades grant RYC-2017-22007 and the European Research Council through the ERC Grant No. 101001561-LISA. A. P. is supported in part by grants CEX2020-001007-S  funded by  MCIN/AEI (Spain)  /10.13039/501100011033 and PID2021-127890NB-I00. The work is further supported by the UK STFC under Grants Nos. ST/L005727/1, ST/P003885/1, ST/V001035/1.
\bibliographystyle{elsarticle-num-names}
\bibliography{draft}

\end{document}